\documentclass[aps,twocolumn,pra,longbibliography,footinbib]{revtex4-2}

\usepackage{amsmath}
\usepackage{amssymb}
\usepackage{booktabs}
\usepackage{array}
\usepackage{mathrsfs}
\usepackage{bm, dsfont}
\usepackage[usenames,dvipsnames]{color}
\usepackage{enumitem}
\usepackage{graphicx}
\usepackage{natbib}
\usepackage[colorlinks=true,linkcolor=blue,citecolor=blue,urlcolor=blue]{hyperref}
\usepackage{cleveref}
\usepackage{hypcap}
\usepackage{physics}
\usepackage{amsthm}

\newcommand{\ii}{\mathrm{i}}
\newcommand{\wt}{\mathrm{wt}}
\newcommand{\Mset}{\mathcal{M}}

\newtheorem{theorem}{Theorem}

\theoremstyle{remark}

\begin{document}

\author{Jef Pauwels}
\author{Nicolas Gisin}
\affiliation{Department of Applied Physics, University of Geneva, Switzerland}
\affiliation{Constructor University, Bremen, Germany}

\title{Tunable Families of Multiqubit Elegant Joint Measurements}

\begin{abstract}
We give a closed-form construction of the $n$-qubit Elegant Joint Measurement (EJM) proposed in [PRL \textbf{136}, 190201 (2026)] and show that it is part of a tunable family of measurements with tetrahedrally arranged Bloch vectors. The construction is based on the interference pattern implied by a single phase polynomial built from the elementary symmetric functions. It realises a regular tetrahedral measurement for every $n$, and the corresponding measurement unitary lies at level $n{+}1$ of the Clifford hierarchy. Starting from this measurement, we ask whether the size of the local tetrahedron---and hence the {single-qubit-versus-rest} entanglement of the basis---can be varied while preserving its symmetry. For every even $n$ the answer is yes, and remarkably the size follows the same one-parameter law that governs the known two-qubit family, interpolating down to a $1$-uniform basis. For $n=3$ the EJM is locally isolated, while for odd $n\ge5$ we do not know an analogous closed-form family. We also give an analogous construction, valid for every $n \geq3$, with square local geometry.
\end{abstract}

\maketitle

\section{Introduction}

Joint measurements are characterized both by the entanglement of their eigenstates and by the geometry of their single-qubit marginals. This raises the questions of how to classify measurements systematically~\cite{Pauwels2025} and how to construct them with prescribed geometry and entanglement~\cite{Pauwels2026PRA}. For pure two-qubit states, the Schmidt decomposition provides a single parameter that tunes the entanglement. For many qubits, special families such as graph and cluster states provide tractable examples~\cite{Briegel2001,Hein2004,Horodecki2009,Guhne2009}. For measurements the situation is more delicate: there is no canonical form even in the bipartite case, few special families are known, and the entanglement can vary from one outcome to another.

A sensible way in is to begin with the most structured measurements and add complexity gradually. The simplest are the \emph{iso-entangled} ones, which we take to mean that their eigenstates form a single local-unitary orbit and therefore share all local-unitary-invariant entanglement properties~\cite{DelSanto2024,Tanaka007,Pimpel2023,Pauwels2026PRA}, and, among these, the ones whose local geometry is most symmetric. Analytically tractable multipartite examples with prescribed regular local geometry are nonetheless scarce. The Elegant Joint Measurement (EJM) is the prime example: the one-qubit marginals of its four basis states point to the vertices of a regular tetrahedron, the most uniform way to place four directions on the Bloch sphere. Its eigenstates are nevertheless only partially entangled, and the measurement can be implemented locally at a low entanglement cost~\cite{Gisin2019,Pauwels2025}. It further belongs to a one-parameter family that interpolates, at fixed tetrahedral symmetry, between the EJM and the maximally entangled Bell basis~\cite{Tavakoli2021}. This combination of symmetry and tunability has made it a recurring ingredient and benchmark in network nonlocality~\cite{Gisin2019,Pozas2022,Gitton2024}, bilocality~\cite{Tavakoli2021,Baeumer2021,Huang2022}, and related tasks~\cite{Ding2024,Patra2025}.

A recent line of work ties the classification of measurements to an operational question: how much entanglement spatially separated parties must share in order to implement a given measurement using only local operations. This entanglement cost of localisation can be read as a measure of the nonlocal complexity of the measurement; it yields a clean classification in principle, though one that is hard to compute in practice~\cite{Pauwels2025}. For two qubits the EJM emerges from this framework as the unique low-complexity measurement with regular tetrahedral symmetry. Beyond two qubits the analysis becomes much harder, but a workable route was opened recently~\cite{MultiqubitEJM}: such measurements can be viewed as orbits of a fixed abelian symmetry group, which reduces their construction to the choice of a single phase polynomial. Related symmetric joint measurements on multiple qubits have been studied from a different construction in Ref.~\cite{Ding2025}. For the present Pauli-orbit family, the Clifford level gives a computable sufficient localisation level, and regular tetrahedral measurements were found by an exhaustive search for three and four qubits. Two features were observed but left unproven: that these measurements should have a closed form valid for any number of qubits $n$, and that their tetrahedron should shrink by a fixed factor with each added qubit~\cite{MultiqubitEJM}. A complementary tunable construction for even $n$, connecting product measurements to the EJM but with a different local geometry, was recently given in Ref.~\cite{He2026}.

Our first result establishes both at once. The interference pattern implied by the phase polynomial
\begin{equation}\label{eq:teaser}
f_{\mathrm{EJM}}=\sum_{k=2}^{n}(-1)^k e_k(z_1,\dots,z_n)
\end{equation}
turns out to define a regular tetrahedral measurement for every $n$, where $e_k(z_1,\dots,z_n)=\sum_{1\le i_1<\cdots<i_k\le n}z_{i_1}\cdots z_{i_k}$ is the $k$th elementary symmetric polynomial. We compute its geometry and complexity in closed form, and show that its measurement unitary sits at level $n{+}1$ of the Clifford hierarchy, certifying localisation at hierarchy level at most $n{+}1$.

A subtler question is whether the full symmetry of the EJM can be preserved while the entanglement entropies are controlled continuously, as it can for two qubits~\cite{Tavakoli2021}. We show that for every \emph{even} number of qubits the answer is yes: perturbing~\eqref{eq:teaser} by a single indicator function generates an exact one-parameter family whose tetrahedron obeys the very same size law $r(\tau)=r_{\mathrm{EJM}}\lvert\cos(\pi\tau/2)\rvert$ and terminates at a $1$-uniform basis~\cite{Scott2004,Arnaud2013,Goyeneche2014}. Within the fixed Pauli-orbit phase-function ansatz, the selected three-qubit EJM is instead locally isolated. For higher odd numbers of qubits we know of no comparable closed form.

To determine which cases admit deformations at all, we examine the local manifold of regular tetrahedral measurements and compute its dimension numerically for small $n$. A square-geometry analogue, valid for every $n\ge3$, is given in Appendix~\ref{sec:square}.

\section{Setup}\label{sec:setup}

Throughout, a measurement means a rank-one projective measurement identified with its orthonormal basis. The tetrahedral measurements we study are convenient to work with for three reasons: each is fixed by a single phase polynomial, with $2^n-1$ real parameters after fixing global phase{, before imposing regular tetrahedral geometry}; it carries a prescribed local symmetry on every qubit by construction; and, as we recall below, that same polynomial provides a computable sufficient localisation level. This section makes these points precise and fixes notation, following the group-orbit description of Refs.~\cite{Pauwels2026PRA,MultiqubitEJM}, to which we refer for all further details.

\begin{figure*}[t]
\centering
\includegraphics[width=\textwidth]{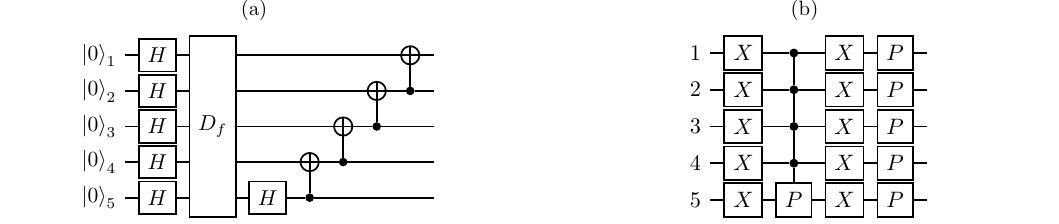}
\caption{{(a)} Normal form for the fiducial state in Eq.~\eqref{eq:normalform}, shown for $n=5$. The first layer is $H^{\otimes n}$, the box is the diagonal phase gate $D_f$, the extra Hadamard is $H_n$, and the CNOT string $S(n)=\prod_{j=2}^n\mathrm{CNOT}_{j\to j-1}$. {From left to right, the CNOTs act in the order $5\to4$, $4\to3$, $3\to2$, $2\to1$. (b) Decomposition of $D_{f_{\mathrm{EJM}}}$ for $n=5$, Eq.~\eqref{eq:ejmgatedecomp}, with $P=\mathrm{diag}(1,\ii)$. Gates act from left to right; the global phase $-\ii$ is omitted.}}
\label{fig:normalform}
\label{fig:ejm-phase-circuit}
\end{figure*}

A tetrahedral basis is the orbit of a fiducial state under the abelian group $G^{(n)}_{\mathrm{tetra}}=\langle Z^{(i)}Z^{(i+1)}\ (i=1,\dots,n-1),X^{\otimes n}\rangle\cong\mathbb{Z}_2^n$, so that each qubit carries a tetrahedral (Pauli-orbit) symmetry. Any such fiducial can be written as
\begin{equation}\label{eq:normalform}
\begin{aligned}
\ket\psi&=S(n)\,H_n\,D_f\,H^{\otimes n}\ket0^{\otimes n},\\
S(n)&=\prod_{j=2}^{n}\mathrm{CNOT}_{j\to j-1}.
\end{aligned}
\end{equation}
This fiducial is realised by the circuit shown in Fig.~\ref{fig:normalform}{(a)}. Our product convention is
\begin{equation*}
S(n)=\mathrm{CNOT}_{2\to1}\mathrm{CNOT}_{3\to2}\cdots\mathrm{CNOT}_{n\to n-1},
\end{equation*}
with the rightmost gate acting first on kets.  Here $H_n$ is the Hadamard on the last qubit and $D_f$ is a diagonal phase gate
\begin{equation}\label{eq:Df}
D_f\ket{\vec z}=\exp\!\Bigl(2\pi\ii\,f(\vec z)/2^m\Bigr)\ket{\vec z},\qquad f(\vec z)=\!\!\sum_{S\subseteq[n]}\!a_S\!\prod_{j\in S}z_j \,
\end{equation}
where $z_j\in\{0,1\}$ label the computational basis states. {At finite phase precision, the} coefficients $a_S\in\{0,1,\dots,2^m-1\}$ are understood modulo $2^m$, with $m$ called the phase precision; unless stated otherwise $m=2$, in which case $D_f\ket{\vec z}=\ii^{f(\vec z)}\ket{\vec z}$. These coefficients define the measurement via the phase polynomial $f$. {For continuous families we allow real coefficients, as in Sec.~\ref{sec:n2}.} The intermediate states $D_f\ket{+}^{\otimes n}$ are diagonal phase states of the same kind that appear in locally entangleable, weighted-graph, and hypergraph-state constructions~\cite{Kruszynska2009,Rossi2013,Guhne2014Hypergraph}; the Pauli orbit in~\eqref{eq:normalform} turns this diagonal-state data into a measurement basis.
The orbit is automatically an orthonormal basis, for every choice of $f$: orthonormality only requires the fiducial to have components of equal modulus along the $2^n$ joint eigenstates of the group, and in the normal form~\eqref{eq:normalform} the diagonal gate changes only phases, never these moduli~\cite{Pauwels2026PRA}.

Geometrically, we call a measurement \emph{regular tetrahedral with Bloch length $r$} when every qubit's reduced Bloch vector $\vec m_j$ has $\lvert\langle X_j\rangle\rvert=\lvert\langle Y_j\rangle\rvert=\lvert\langle Z_j\rangle\rvert$ and $\lVert\vec m_j\rVert=r$; the orbit then produces a congruent tetrahedron on every qubit, with $r=0$ included as the limiting case.

A further advantage of this description is that the cost of localising the measurement---the entanglement two parties must share to implement it with local operations---is easy to bound: it is upper bounded by the Clifford-hierarchy level of the diagonal gate $D_f$~\cite{Pauwels2026PRA}, the only non-Clifford resource in the normal-form circuit, and for a phase polynomial this level can be read directly off the coefficients~\cite{Gottesman2017},
\begin{equation}\label{eq:level}
k=\max_{S:\,a_S\ne0}\bigl[(m-\nu_2(a_S)-1)+\lvert S\rvert\bigr],
\end{equation}
with $\nu_2$ the $2$-adic valuation.

Both the local geometry and this implementation complexity are therefore read off from the same object $f$. In this language the two-qubit EJM is simply $f=z_1z_2$ at level $3$, the unique regular tetrahedral, efficiently localisable two-qubit measurement~\cite{MultiqubitEJM}. Beyond two qubits it is no longer unique: an exhaustive search~\cite{MultiqubitEJM} found regular tetrahedral EJMs for $n=3$ and $n=4$ (at levels $4$ and $5$) falling into several locally inequivalent classes, a first sign of the richer entanglement structure of multiqubit measurements. The search became intractable for $n\ge5$, and it was conjectured that such measurements persist for every $n$, at Clifford level $n{+}1$ and Bloch length $\sqrt3/2^{\,n-1}$~\cite{MultiqubitEJM}. The next theorem establishes the Bloch length exactly and fixes the Clifford level at $n{+}1$, which upper-bounds the localisation cost.

\section{Closed-form \texorpdfstring{$n$}{n}-qubit EJM}\label{sec:ejm}

Starting from the two-qubit form $f=z_1z_2$, the natural way to keep its symmetry as qubits are added is to stay within the symmetric polynomials, and the right combination is~\eqref{eq:teaser}.

\begin{theorem}\label{thm:ejm}
For every $n\ge2$ the precision-$2$ polynomial $f_{\mathrm{EJM}}=\sum_{k=2}^{n}(-1)^k e_k\!\pmod4$ defines, through Eq.~\eqref{eq:normalform}, a regular tetrahedral $n$-qubit measurement with Bloch vectors
\begin{equation}\label{eq:ejmbloch}
\vec m_j=\tfrac{1}{2^{n-1}}(1,1,1)\ (j<n),\quad \vec m_n=\tfrac{1}{2^{n-1}}(1,-1,1),
\end{equation}
hence $r_{\mathrm{EJM}}=\sqrt3/2^{\,n-1}$, and its measurement unitary sits exactly at Clifford level $n{+}1$.
\end{theorem}

Two short observations make the theorem transparent; the full Bloch computation is in Appendix~\ref{app:proof}.

First, $f_{\mathrm{EJM}}$ depends only on the Hamming weight $w=\wt(\vec z)$, the number of $1$'s in $\vec z$. Indeed each $e_k$ is symmetric, and on binary inputs the monomial $z_{i_1}\cdots z_{i_k}$ equals $1$ exactly when all $k$ of its indices fall among the $w$ ones; counting these subsets gives $e_k(\vec z)=\binom{w}{k}$, a function of $w$ alone. The alternating sum then follows from the binomial theorem\footnote{The binomial theorem states that $(1+x)^n = \sum_{k=0}^n \binom{n}{k} x^{k}$. Setting $x=-1$ gives $(1-1)^n = \sum_{k=0}^n \binom{n}{k} (-1)^k$. } $\sum_{k=0}^w(-1)^k\binom{w}{k}=(1-1)^w$, after separating off the $k=0$ and $k=1$ terms,
\begin{equation}\label{eq:fwt}
f_{\mathrm{EJM}}(\vec z)=(1-1)^w-1+w=\begin{cases}0,&w\le1,\\ w-1\!\!\pmod4,&w\ge2,\end{cases}
\end{equation}
so $D_f$ multiplies each weight-$w$ string by $\ii^{\,w-1}$ for $w\ge2$. Explicitly $f_{\mathrm{EJM}}=z_1z_2$ for $n=2$, ${e_2-e_3}\mathrel{{\equiv}}{e_2+3e_3}\mathrel{{=}}z_1z_2+z_1z_3+z_2z_3+3z_1z_2z_3$ for $n=3$, and ${e_2-e_3+e_4}\mathrel{{\equiv}}e_2+3e_3+e_4$ for $n=4${, where the equivalences are modulo $4$ and $-1\equiv3\pmod4$}; these reproduce the symmetric classes found numerically in~\cite{MultiqubitEJM}.

Second, the EJM gate is a tensor product of single-qubit gates, corrected on a single string. Indeed, by~\eqref{eq:fwt} it multiplies every string of weight $w\ge1$ by $\ii^{\,w-1}=-\ii\cdot\ii^{\,w}$ and the all-zero string by $1$. The product of single-qubit phase gates $\mathrm{diag}(1,\ii)^{\otimes n}$ produces exactly the phase $\ii^{\,w}$---one factor $\ii$ for each qubit in state $\ket1$---so, as an operator identity,
\begin{equation}\label{eq:rankone}
D_{f_{\mathrm{EJM}}}=-\ii\,\mathrm{diag}(1,\ii)^{\otimes n}+(1+\ii)\bigl(\ket0\!\bra0\bigr)^{\otimes n},
\end{equation}
where the rank-one term repairs the phase of the all-zero string ($-\ii+1+\ii=1$). A product gate creates no entanglement, so that term is the only source of it: the pre-CNOT fiducial $\ket{\phi_n}=H_nD_fH^{\otimes n}\ket0^{\otimes n}$ collapses into a superposition of just two product states,
\begin{equation}\label{eq:phidecomp}
\ket{\phi_n}=\frac{(1+\ii)\ket{A_n}+(1-\ii)\ket{B_n}}{\sqrt{2^{\,n+1}}},
\end{equation}
with $\ket{A_n}=\ket0^{\otimes n-1}(\ket0+\ket1)$ and $\ket{B_n}=(\ket0+\ii\ket1)^{\otimes n-1}(\ket0-\ii\ket1)$: the branch $\ket{B_n}$ comes from the product gate acting on the uniform superposition, the branch $\ket{A_n}$ from the rank-one correction, each pushed through the final Hadamard $H_n$. Each single-qubit marginal is then fixed by the interference of these two branches, which gives the Bloch vectors~\eqref{eq:ejmbloch}. Both steps---the collapse to~\eqref{eq:phidecomp} and the resulting marginals---are carried out in Appendix~\ref{app:proof}.

\begin{figure}[b]
\centering
\includegraphics{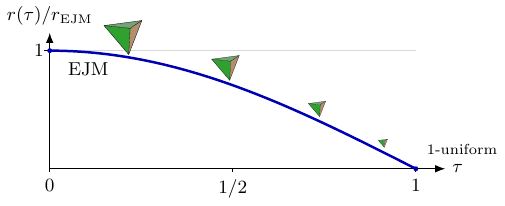}
\caption{Bloch-length of the even-$n$ family. The local tetrahedra keep their orientation and shrink by the common factor $\lvert\cos(\pi\tau/2)\rvert$, reaching a $1$-uniform endpoint at $\tau=1$, where all one-qubit marginals are $I/2$.}
\label{fig:bloch-length}
\end{figure}

\begingroup
For implementation, write $P=\mathrm{diag}(1,\ii)$. The gate~\eqref{eq:rankone} also has the circuit decomposition
\begin{equation}\label{eq:ejmgatedecomp}
D_{f_{\mathrm{EJM}}}=-\ii\,P^{\otimes n}X^{\otimes n}C^{n-1}(P)X^{\otimes n}.
\end{equation}
Here $C^{n-1}(P)$ applies $P$ to the last qubit conditioned on the other $n-1$ qubits being in $\ket1$. Conjugating by $X^{\otimes n}$ supplies the phase correction on $\ket{0^n}$; Fig.~\ref{fig:ejm-phase-circuit}(b) shows this decomposition for $n=5$.
\endgroup

Applying $S(n)$ gives the fiducial $\ket{\psi_n}$, whose Pauli orbit is the full measurement; Fig.~\ref{fig:explicit-ejm-bases} gives the resulting basis states explicitly for $n=2,3,4$. {For example, in the order $\ket{000},\ket{001},\ldots,\ket{111}$, the three-qubit fiducial is}
\begingroup
\begin{equation}\label{eq:ejm3fiducial}
\ket{\psi_3}=\frac14\bigl(2,1-\ii,-1+\ii,1-\ii,1+\ii,1+\ii,1+\ii,0\bigr)^T.
\end{equation}
Its eight orthonormal basis states are
\begin{equation}\label{eq:ejm3orbit}
(Z_1Z_2)^a(Z_2Z_3)^b(X_1X_2X_3)^c\ket{\psi_3},\quad a,b,c\in\{0,1\}.
\end{equation}
These are the columns of the $n=3$ matrix in Fig.~\ref{fig:explicit-ejm-bases}, normalized by $1/4$, up to their ordering.
\endgroup

The Clifford level follows directly from~\eqref{eq:level}: every monomial of $f_{\mathrm{EJM}}$ has an odd coefficient, so its contribution is $1+\lvert S\rvert$, maximised by the unique degree-$n$ term $z_1\cdots z_n$. Hence the level is exactly $n{+}1$, non-Clifford for all $n$ and growing linearly with the qubit number. The conjugate polynomial $\sum(-1)^{k+1}e_k$ gives the mirror-image (complex-conjugate) EJM within the same phase-polynomial normal form.

Theorem~\ref{thm:ejm} thus turns the two empirical patterns of~\cite{MultiqubitEJM}---a closed form and the $\sqrt3/2^{n-1}$ scaling---into a single statement, proved for all $n$.

\section{Tuning the entanglement}\label{sec:families}

We now keep the local symmetry of the measurement fixed and ask how much the size of its tetrahedron can be made to vary. The Bloch length $r$ directly quantifies the one-qubit-versus-rest entanglement entropies within this family: every single-qubit reduced state has eigenvalues $(1\pm r)/2$, so the entanglement entropy between any one qubit and the rest is the binary entropy $h\bigl(\tfrac{1+r}{2}\bigr)$, with $h(x)=-x\log_2x-(1-x)\log_2(1-x)$---the same for every eigenstate and every qubit, since all eigenstates are related by local unitaries. Shrinking the tetrahedron therefore increases the entanglement across every single-qubit cut, up to the maximum at $r=0$, where each qubit is maximally entangled with the rest.

\subsection{The two-qubit family}\label{sec:n2}

For two qubits such a construction is known: Ref.~\cite{Tavakoli2021} constructed a one-parameter family of iso-entangled bases interpolating between the EJM and the Bell basis while keeping the tetrahedral symmetry, and Ref.~\cite{Pauwels2026PRA} recovered it within the present group-orbit framework as a line of fiducial phases. In the phase-polynomial language the family becomes a straight segment between two symmetric polynomials,
\begin{equation}\label{eq:cyril}
f_\tau=(1-\tau)\,e_2+\tau\,e_1,\qquad \tau\in[0,1],
\end{equation}
where $f_\tau$ now takes real values, keeping the phase convention of~\eqref{eq:Df} with the precision-$2$ normalisation: explicitly, $D_{f_\tau}=\mathrm{diag}(1,\ii^\tau,\ii^\tau,\ii^{\tau+1})$ with $\ii^\tau=e^{\ii\pi\tau/2}$. At $\tau=0$ this is the EJM, $f=e_2=z_1z_2$. At $\tau=1$ it is $f=e_1=z_1+z_2$, whose diagonal gate is a product of single-qubit phase gates: the orbit is the Bell basis, and the tetrahedron has collapsed to a point, $r=0$. In between, a one-line Bloch computation gives
\begin{equation*}
r(\tau)=\tfrac{\sqrt3}{2}\,\lvert\cos(\pi\tau/2)\rvert,
\end{equation*}
so as $\tau$ runs from $0$ to $1$ the two local tetrahedra keep their orientation and contract smoothly to a point, sweeping the iso-entangled bases from the partially entangled EJM to the maximally entangled Bell basis. (The parameter $\tau$ corresponds to the interpolation angle $\theta=\pi\tau/2$ in the parametrisation of Ref.~\cite{Tavakoli2021}.) We now show that the same one-parameter deformation exists for every even $n$.

\begingroup
\label{para:he-comparison}The three-parameter construction of He \emph{et al.}~\cite{He2025Generalized} also has regular tetrahedral marginals, but allows their orientation to vary relative to the fixed Pauli axes used here.
\endgroup

\subsection{An exact even-\texorpdfstring{$n$}{n} family}\label{sec:even}

To generalise this family, first rewrite it in a way that no longer refers to two qubits. On binary inputs, $e_1-e_2=1-(1-z_1)(1-z_2)$ is the indicator function of the strings other than the all-zero string, so~\eqref{eq:cyril} is $f_\tau=f_{\mathrm{EJM}}+\tau\,\varphi_2$ with $\varphi_2$ that indicator: the deformation simply multiplies the amplitude of every string except the all-zero one by the common phase $e^{\ii\pi\tau/2}$. The naive $n$-qubit guess, i.e. apply this common phase to every string except the all-zero one, fails for $n\ge3$: it already breaks the regular tetrahedral symmetry. The correct generalisation keeps the same form but shrinks the support. For even $n=2t$, let $\varphi$ be the indicator function of the strings in which the even-indexed qubits $2,4,\dots,n-2$ all vanish, again excluding the all-zero string,
\begin{equation}\label{eq:phi}
\varphi(\vec z)=\prod_{k=1}^{t-1}(1-z_{2k})-\prod_{j=1}^{n}(1-z_j),
\end{equation}
which for $n=2$ reduces to $\varphi_2$. We then have the following exact result.

\begin{theorem}\label{thm:even}
For every even $n=2t$, the real-valued phase function
\begin{equation}\label{eq:evenfamily}
f_\tau=f_{\mathrm{EJM}}+\tau\,\varphi,\qquad \tau\in[0,1],
\end{equation}
defines a regular tetrahedral measurement with Bloch length
\begin{equation}\label{eq:blocheven}
r(\tau)=r_{\mathrm{EJM}}\,\lvert\cos(\pi\tau/2)\rvert=\frac{\sqrt3}{2^{\,n-1}}\lvert\cos(\pi\tau/2)\rvert .
\end{equation}
At $\tau=0$ it is the EJM; at $\tau=1$ the tetrahedron has collapsed to a point, $r=0$, and the basis is $1$-uniform.
\end{theorem}

The proof, given in Appendix~\ref{app:evenproof}, shows more strongly that every Bloch component is multiplied by the same factor $\cos(\pi\tau/2)$: the deformation does not tilt or distort the tetrahedra; it only changes their common size. Note that~\eqref{eq:evenfamily} is the straight interpolation $f_\tau=(1-\tau)f_{\mathrm{EJM}}+\tau g$ between the EJM and the endpoint polynomial $g=f_{\mathrm{EJM}}+\varphi$, in direct analogy with~\eqref{eq:cyril}. The size law~\eqref{eq:blocheven}, shown in Fig.~\ref{fig:bloch-length}, is identical to the two-qubit law, rescaled by $r_{\mathrm{EJM}}$.

\paragraph*{Why a cosine, and why even $n$.}
The mechanism builds directly on the two-branch structure of Sec.~\ref{sec:ejm}: because the EJM gate is a product gate plus a rank-one correction, Eq.~\eqref{eq:rankone}, its fiducial is the superposition~\eqref{eq:phidecomp} of only two product branches, and all Bloch vectors come from the interference between $\ket{A_n}$ and $\ket{B_n}$ (Appendix~\ref{app:proof}).

The deformation preserves this product structure. It multiplies the amplitude of every string with $z_2=z_4=\cdots=z_{n-2}=0$ by $e^{\ii\theta}$, $\theta=\pi\tau/2$ (except the all-zero string). This condition acts qubit by qubit, so the part of the product branch $\ket{B_n}$ that satisfies it is itself a product state: $\ket{M_n}$, obtained from $\ket{B_n}$ by resetting the qubits in $E=\{2,4,\dots,n-2\}$ to $\ket0$. Keeping track of the excluded all-zero string (Appendix~\ref{app:evenproof}), the deformed fiducial is exactly a superposition of three product branches,
\begin{equation}\label{eq:threebranch}
\ket{\phi^\tau_n}=\frac{c_A\ket{A_n}+c_B\ket{B_n}+c_M\ket{M_n}}{\sqrt{2^{\,n+1}}},
\end{equation}
with $c_A=1+\ii e^{\ii\theta}$, $c_B=1-\ii$ and $c_M=(1-\ii)(e^{\ii\theta}-1)$; at $\tau=0$ it reduces to~\eqref{eq:phidecomp}. Every Bloch component is again a sum of pairwise interference terms. The calculation in Appendix~\ref{app:evenproof} shows that the terms involving $\ket{M_n}$ either vanish or exactly equal the corresponding EJM terms, and their coefficients combine into the common factor $\cos\theta$ for every Bloch component. (For $n=2$ no qubit is reset, $\ket{M_n}=\ket{B_n}$, and the deformation simply rotates the relative phase of the two branches.)

The alternating pattern in~\eqref{eq:phi} is forced by this mechanism: the interference terms cancel in the required pattern only if no two of the zeroed qubits are adjacent, every adjacent pair of qubits among $1,\dots,n-1$ contains one, and the boundary qubits $1,n-1,n$ remain outside the zeroed set. These conditions force $\{2,4,\dots,n-2\}$ and are possible only for even $n$ (see the closing remark of Appendix~\ref{app:evenproof}). An exhaustive numerical scan over all supports obtained by requiring an arbitrary subset of qubits to vanish confirms, for $n\le8$, that no other choice obeys the cosine law, and that for odd $n$ none does.

\paragraph*{The endpoint.}
At $\tau=1$ the tetrahedron has collapsed and every single-qubit marginal state is exactly $I/2$: the basis is $1$-uniform, with each qubit maximally entangled with the rest. The endpoint polynomial itself takes a simple explicit form. Splitting the qubit indices into $E=\{2,4,\dots,n-2\}$ and $F=[n]\setminus E$, one finds
\begin{equation}\label{eq:endpoint}
g=e_1(z_F)+\sum_{k=2}^{{t}-1}(-1)^k e_k(z_E)\pmod4,
\end{equation}
i.e.\ a \emph{smaller} EJM fiducial: the $({t}-1)$-qubit polynomial $f_{\mathrm{EJM}}$ acting on the even-indexed qubits, together with single-qubit phase gates on the rest. Its Clifford level is $\max(2,n/2)$. The endpoint fiducial is a stabilizer state only for $n=2,4$, and is nonstabilizer from $n=6$ onwards. The simplest case $n=4$ is fully explicit: there $g=z_1+z_3+z_4$ and the endpoint fiducial factorises into two Bell pairs, on qubits $(1,2)$ and $(3,4)$.

\begingroup
\label{sec:multipartite}The three-tangle and pairwise concurrences of the three-qubit EJM were studied in Ref.~\cite{MultiqubitEJM}; Appendix~\ref{app:multipartite} recalls these results and gives the four-qubit genuine multipartite concurrence. For four qubits, the largest tetrahedra in this family ($\tau=0$) already yield a genuine multipartite concurrence~\cite{Ma2011GME} $C_{\mathrm{GME}}=\sqrt{61}/8\simeq0.976$, close to its maximum possible value of $1$, whereas the Bell-pair endpoint has vanishing genuine multipartite concurrence despite maximal single-qubit entropies.
\endgroup

\paragraph*{Complexity along the way.}
For dyadic $\tau=\ell/2^{m-2}$ written at minimal precision $m$, the family stays within that finite phase precision, and~\eqref{eq:level} gives level $n+1-\nu_2(1-\tau)$ for $\tau\ne1$, where $\nu_2$ is the $2$-adic valuation of the rational number $1-\tau$. The level drops to $\max(2,n/2)$ at the endpoint. For non-dyadic $\tau$ there is no exact finite phase precision, hence no finite Clifford-hierarchy level in the sense of~\eqref{eq:level}. This generalizes the observations of \cite{Pauwels2026PRA} for the two-qubit family.

\subsection{Odd \texorpdfstring{$n$}{n}}\label{sec:odd}

The construction has no odd-$n$ analogue of the same form: as noted above, no support of the type~\eqref{eq:phi} reproduces the cosine law for odd $n$ (checked exhaustively for $n=3,5,7$). For $n=3$ this is not just a limitation of our construction: the selected EJM is locally isolated within our ansatz (Sec.~\ref{sec:manifold}). This does not contradict the three-qubit tetrahedral-symmetry families of Ref.~\cite{Ding2025}, which lie outside the present regular tetrahedral class. For $n=5$ the local manifold has positive dimension, so tunable families are not ruled out---only this particular mechanism fails---but we currently know no closed-form odd-$n$ family with the universal cosine law. The exact family of Theorem~\ref{thm:even} is thus specific to even numbers of qubits.

\section{The solution manifold}\label{sec:manifold}

How much freedom is there to deform the EJM while keeping its regular tetrahedral symmetry? A simple count of parameters and constraints answers this locally. Let $\Mset_n$ denote the set of phase polynomials whose measurement is regular tetrahedral. A phase polynomial has $2^n-1$ real coefficients $a_S$. Regularity imposes $3n-1$ polynomial equations on them: on each qubit the three squared Bloch components must be equal ($2n$ equations), and the Bloch lengths must agree across qubits ($n-1$ more). Wherever the Jacobian $J$ of these equations has full row rank at the EJM---possible only once $2^n-1\ge3n-1$, i.e.\ $n\ge4$---the implicit function theorem gives
\begin{equation}\label{eq:dim}
\dim\Mset_n=2^n-3n .
\end{equation}
Computing $\operatorname{rank}J$ numerically settles the first cases (Table~\ref{tab:dim}). For $n=2$ only two of the five equations are independent, so exactly one direction of deformation survives: this is the family of Sec.~\ref{sec:n2}. For $n=3$ the rank is seven, as large as the number of coefficients, so $\dim\Mset_3=0$: the three-qubit EJM is an isolated point, and no continuous family passes through it. For $n=4,5,6$ the Jacobian has full row rank and the solution space is large. The exact even-$n$ line of Theorem~\ref{thm:even} is a single curve inside it; for four qubits we also found, numerically, a second solution with a larger tetrahedron, lying on its own straight line to a different $1$-uniform endpoint (Appendix~\ref{app:159}).

\begin{table}[t]
\centering
\renewcommand{\arraystretch}{1.15}
\begin{tabular}{lccccc}
\toprule
$n$ & $2$ & $3$ & $4$ & $5$ & $6$\\
\midrule
coefficients $2^n-1$ & $3$ & $7$ & $15$ & $31$ & $63$\\
constraints $3n-1$ & $5$ & $8$ & $11$ & $14$ & $17$\\
$\operatorname{rank}J$ & $2$ & $7$ & $11$ & $14$ & $17$\\
$\dim\Mset_n$ & $1$ & $0$ & $4$ & $17$ & $46$\\
\bottomrule
\end{tabular}
\caption{Numerical Jacobian ranks and resulting local dimensions of the regular tetrahedral manifold at the EJM. The one-dimensional $n=2$ case is the tunable family; the zero-dimensional $n=3$ case is the isolated EJM.}
\label{tab:dim}
\end{table}

\section{Discussion}\label{sec:discussion}

We presented a simple closed-form family of EJMs for any number of qubits $n$, proving the main conjecture of Ref.~\cite{MultiqubitEJM}. For even $n$ we further present a one-parameter family of iso-entangled, regular tetrahedral bases that smoothly connect these EJMs to a $1$-uniform basis, allowing to control the degree of entanglement while keeping the local symmetry fixed. For the two-qubit case, this reduces to the known construction of Ref.~\cite{Tavakoli2021}.

The most natural application for these measurements is in networks, where a multiqubit EJM can sit at a central node and the parameter $\tau$ directly controls the entanglement entropy. The behavior family in star and fully connected networks remains open; related three-qubit symmetric measurements in a star network were studied in Ref.~\cite{Ding2025}. Understanding how tuning $\tau$ changes the resulting correlations and their nonlocality, as was done in Ref.~\cite{Tavakoli2021} for the bilocality network, is an interesting open question. On the structural side, for $n=5$ the local manifold already has large dimension, but we have not yet found any closed-form family inside it; and the global relation between the even-$n$ family and the second four-qubit line of Appendix~\ref{app:159} is unknown. Finally, the square construction suggests a broader class of measurements with regular polygonal local geometry, which it would be interesting to classify systematically. {These commuting Pauli-orbit families also fit into the broader classification of local encodings in Ref.~\cite{PauwelsGuhne2026}, which shows that not every multiqubit state can be extended to an iso-entangled measurement basis.}

\begin{acknowledgments}
We acknowledge support from the Swiss National Science Foundation (NCCR-SwissMAP). We used OpenAI Codex (GPT-5.4 High) and Anthropic Claude Fable 5 to help guess the explicit form of the $n$-qubit family, for editorial assistance and to polish the code. All output was reviewed and edited by the authors.

The numerical checks reported in this paper are available, with instructions, in Ref.~\cite{EJMCode}.
\end{acknowledgments}

\bibliography{refs}
\appendix

\section{Explicit EJM basis matrices}\label{sm:explicit-bases}
\begin{figure*}[t]
\centering
\includegraphics[width=0.98\textwidth]{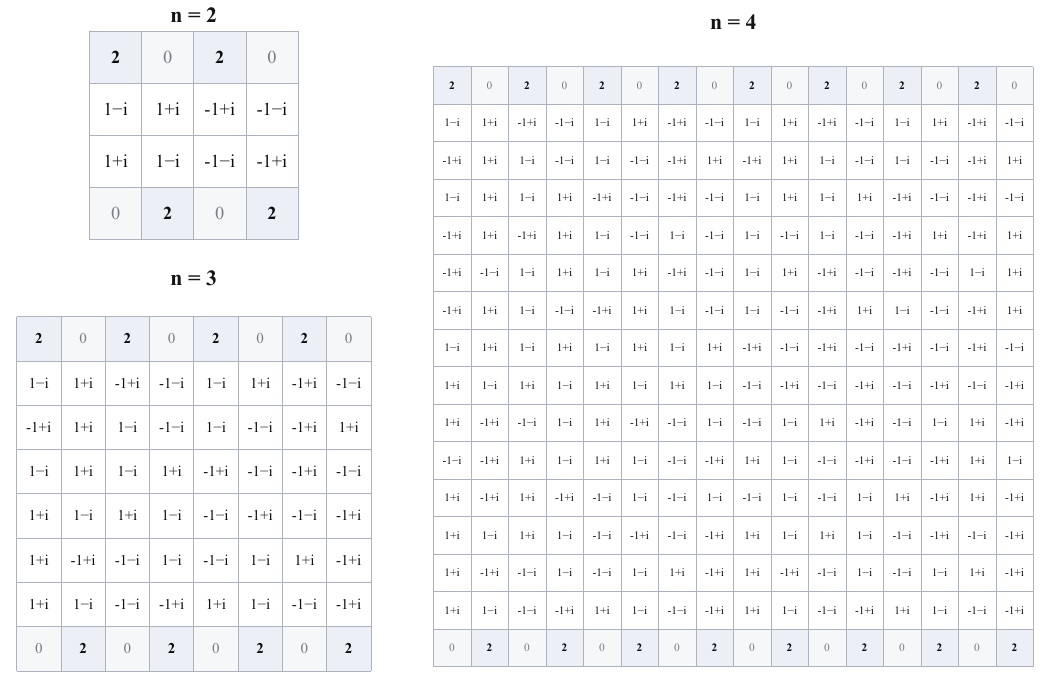}
\caption{Explicit EJM basis matrices for $n=2,3,4$. Their columns are the Pauli-orbit basis states in the computational basis. We write $M_n=2^{-(n+1)/2}A_n$; the displayed numbers are the entries of $A_n$.}
\label{fig:explicit-ejm-bases}
\end{figure*}

Figure~\ref{fig:explicit-ejm-bases} gives the EJM basis matrices for $n=2,3,4$.

\section{Bloch vectors of the EJM}\label{app:proof}

In this appendix, we first derive the form~\eqref{eq:phidecomp} and then compute the marginals of Theorem~\ref{thm:ejm}.

\emph{The two-branch form~\eqref{eq:phidecomp}.} Applying the Hadamards to $\ket0^{\otimes n}$ gives the uniform superposition $2^{-n/2}\sum_{\vec z}\ket{\vec z}$, on which we act with the gate~\eqref{eq:rankone}. The product gate turns the uniform superposition into a product state, $\mathrm{diag}(1,\ii)^{\otimes n}\,2^{-n/2}\sum_{\vec z}\ket{\vec z}=2^{-n/2}(\ket0+\ii\ket1)^{\otimes n}$, each qubit picking up its phase independently, while the rank-one term contributes $(1+\ii)\,2^{-n/2}\ket0^{\otimes n}$. Hence
\[
D_fH^{\otimes n}\ket0^{\otimes n}=2^{-n/2}\bigl[(1+\ii)\ket0^{\otimes n}-\ii\,(\ket0+\ii\ket1)^{\otimes n}\bigr].
\]
The final Hadamard on qubit $n$ sends $\ket0_n\mapsto\tfrac1{\sqrt2}(\ket0+\ket1)$ and $(\ket0+\ii\ket1)_n\mapsto\tfrac{1+\ii}{\sqrt2}(\ket0-\ii\ket1)$, hence $\ket0^{\otimes n}\mapsto\tfrac1{\sqrt2}\ket{A_n}$ and $(\ket0+\ii\ket1)^{\otimes n}\mapsto\tfrac{1+\ii}{\sqrt2}\ket{B_n}$. Substituting and using $-\ii(1+\ii)=1-\ii$ yields the two-term form~\eqref{eq:phidecomp}.

\emph{The marginals.} Because $\ket{\psi_n}=S(n)\ket{\phi_n}$, a single-qubit Pauli average is
\[
\langle P_j\rangle_{\psi_n}
=\bra{\phi_n}S(n)^\dagger P_jS(n)\ket{\phi_n},
\qquad P\in\{X,Y,Z\}.
\]
Thus we need the Heisenberg relations for the CNOT string. The two basic ones are
\begin{align}
S^\dagger Z_j S&=Z_jZ_{j+1}\cdots Z_n,\\
S^\dagger X_1 S&=X_1,\qquad S^\dagger X_j S=X_{j-1}X_j\quad (j\ge2).
\end{align}
From $Y_j=\ii X_jZ_j$,
\begin{equation}
S^\dagger Y_jS=\ii\,(S^\dagger X_jS)(S^\dagger Z_jS),
\end{equation}
which gives $Y_1Z_2\cdots Z_n${ for $j=1$}, $X_{j-1}Y_jZ_{j+1}\cdots Z_n$ for $2\le j\le n-1$, and $X_{n-1}Y_n$ for $j=n$.

Writing $\ket{A_n}={\lvert}\alpha{\rangle}^{\otimes n-1}{\lvert}\gamma{\rangle}$ and $\ket{B_n}={\lvert}\beta{\rangle}^{\otimes n-1}{\lvert}\delta{\rangle}$ with {unnormalized kets} ${\lvert}\alpha{\rangle}=\ket0$, ${\lvert}\beta{\rangle}=\ket0+\ii\ket1$, ${\lvert}\gamma{\rangle}=\ket0+\ket1$, ${\lvert}\delta{\rangle}=\ket0-\ii\ket1$, every matrix element factorises over qubits into single-qubit brackets, which we collect below (Appendix~\ref{app:evenproof} reuses them):
\begin{equation}\label{eq:brackets}
\begin{array}{c|cccc}
 & I & X & Y & Z\\
\hline
\bra\alpha\,\cdot\,\ket\alpha & 1 & 0 & 0 & 1\\
\bra\beta\,\cdot\,\ket\beta & 2 & 0 & 2 & 0\\
\bra\gamma\,\cdot\,\ket\gamma & 2 & 2 & 0 & 0\\
\bra\delta\,\cdot\,\ket\delta & 2 & 0 & -2 & 0\\
\bra\alpha\,\cdot\,\ket\beta & 1 & \ii & 1 & 1\\
\bra\gamma\,\cdot\,\ket\delta & 1-\ii & 1-\ii & -1+\ii & 1+\ii
\end{array}
\end{equation}
Write $O_{j,P}=S^\dagger P_jS$ for the pulled-back Pauli strings listed above. The diagonal branch terms vanish,
\[
\bra{A_n}O_{j,P}\ket{A_n}
=\bra{B_n}O_{j,P}\ket{B_n}=0,
\]
because every $O_{j,P}$ contains at least one factor with a vanishing bracket in~\eqref{eq:brackets}: an $X$ or $Z$ acting on ${\lvert}\beta{\rangle}$ or ${\lvert}\delta{\rangle}$ kills the $\ket{B_n}$ term, and an $X$ or $Y$ acting on ${\lvert}\alpha{\rangle}$, or the final $Z_n$ acting on ${\lvert}\gamma{\rangle}$, kills the $\ket{A_n}$ term. So only the interference between the two product branches matters. Let
\[
M_{j,P}:=\bra{A_n}O_{j,P}\ket{B_n},\qquad P\in\{X,Y,Z\}.
\]
From the rules above, the $M_{j,P}$ are
\[
\begin{array}{c|ccc}
&M_{j,X}&M_{j,Y}&M_{j,Z}\\
\hline
j=1&1+\ii&1+\ii&1+\ii\\
2\le j\le n-1&-1+\ii&-1+\ii&1+\ii\\
j=n&1+\ii&-1-\ii&1+\ii
\end{array}
\]
(the middle row is absent when $n=2$). Finally, because
\begin{align}
\langle P_j\rangle_{\psi_n}
&=2^{-(n+1)}\bigl[(1-\ii)^2M_{j,P}+(1+\ii)^2M_{j,P}^*\bigr]\notag\\
&=2^{1-n}\operatorname{Im}M_{j,P},
\end{align}
we immediately get $\langle X_j\rangle=\langle Z_j\rangle=2^{1-n}$ for all $j$, $\langle Y_j\rangle=2^{1-n}$ for $j<n$ and $\langle Y_n\rangle=-2^{1-n}$. This is~\eqref{eq:ejmbloch}, with $\lVert\vec m_j\rVert=\sqrt3/2^{\,n-1}$. \hfill$\qed$

\section{Proof of Theorem~\ref{thm:even}}\label{app:evenproof}

We prove the stronger statement $\vec m_j(\tau)=\cos(\tfrac{\pi\tau}{2})\,\vec m_j(0)$ for every qubit $j$, which contains~\eqref{eq:blocheven}. Everything is computed with the tools of Appendix~\ref{app:proof}: the branch states $\ket{A_n}$, $\ket{B_n}$, the pulled-back Pauli strings $O_{j,P}$, the bracket table~\eqref{eq:brackets}, and the interference terms $M_{j,P}$. Throughout, $\theta=\pi\tau/2$, $E=\{2,4,\dots,n-2\}$ denotes the even-indexed qubits and $F$ the remaining $t+1$ qubits.

\paragraph*{Step 1: the three-branch form~\eqref{eq:threebranch}.}
Since $\varphi$ is the indicator of its support $S$, the diagonal gate is the EJM gate followed by a single conditional phase,
\begin{equation}\label{eq:gatedecomp}
D_{f_\tau}=D_{f_{\mathrm{EJM}}}\bigl[(I-\Pi_S)+e^{\ii\theta}\Pi_S\bigr],
\end{equation}
with $\Pi_S$ the projector onto the $2^{{t}+1}-1$ strings in $S$. Hence the deformed fiducial $\ket{\phi^\tau_n}=H_nD_{f_\tau}H^{\otimes n}\ket0^{\otimes n}$ is the EJM fiducial plus a correction from the support,
\[
\ket{\phi^\tau_n}=\ket{\phi_n}+\bigl(e^{\ii\theta}-1\bigr)\,H_nD_{f_{\mathrm{EJM}}}\Pi_S\ket{+}^{\otimes n}.
\]
Because the all-zero string is excluded from $S$, the rank-one term of~\eqref{eq:rankone} gives zero on $\Pi_S\ket{+}^{\otimes n}$, so only the product gate acts there: with $\Pi_S\ket{+}^{\otimes n}=2^{-n/2}\bigl[\ket0_E\otimes(\ket0+\ket1)^{\otimes({t}+1)}_F-\ket0^{\otimes n}\bigr]$,
\[
{
\begin{aligned}
D_{f_{\mathrm{EJM}}}\Pi_S\ket{+}^{\otimes n}
&=-\ii\,2^{-n/2}\bigl[\ket0_E\otimes
  (\ket0+\ii\ket1)^{\otimes({t}+1)}_F\\
&\hspace{4.5em}-\ket0^{\otimes n}\bigr].
\end{aligned}}
\]
The final Hadamard acts exactly as in Appendix~\ref{app:proof} ($\ket0_n\mapsto{\lvert}\gamma{\rangle}/\sqrt2$, $(\ket0+\ii\ket1)_n\mapsto(1+\ii)\,{\lvert}\delta{\rangle}/\sqrt2$), turning the bracket into $\tfrac{1+\ii}{\sqrt2}\ket{M_n}-\tfrac1{\sqrt2}\ket{A_n}$ with
\[
\ket{M_n}={\lvert}\mu_1{\rangle}\otimes\cdots\otimes{\lvert}\mu_{n-1}{\rangle}\otimes{\lvert}\delta{\rangle},\qquad
{\lvert}\mu_j{\rangle}=\begin{cases}{\lvert}\alpha{\rangle},& j\in E,\\ {\lvert}\beta{\rangle},& j\notin E,\end{cases}
\]
i.e.\ $\ket{B_n}$ with the qubits in $E$ reset to $\ket0$, as stated in the main text. Collecting coefficients (using $-\ii(1+\ii)=1-\ii$) gives exactly the three-branch form~\eqref{eq:threebranch}, and setting $\tau=0$ recovers~\eqref{eq:phidecomp}.

\paragraph*{Step 2: only three interference terms survive.}
Exactly as in Appendix~\ref{app:proof}, $\langle P_j\rangle_\tau=\bra{\phi^\tau_n}O_{j,P}\ket{\phi^\tau_n}$, and expanding~\eqref{eq:threebranch} gives nine terms. The three diagonal ones vanish by the same bracket argument as before, since $\ket{M_n}$ is again a product of ${\lvert}\alpha{\rangle}$, ${\lvert}\beta{\rangle}$ and ${\lvert}\delta{\rangle}$ factors. What remains is
\begin{equation}\label{eq:threeterms}
\langle P_j\rangle_\tau=2^{-n}\,\mathrm{Re}\bigl[\bar c_Ac_B\,M_{j,P}+\bar c_Ac_M\,N_{j,P}+\bar c_Bc_M\,K_{j,P}\bigr],
\end{equation}
with the known EJM term $M_{j,P}$ and two new columns,
\[
N_{j,P}=\bra{A_n}O_{j,P}\ket{M_n},\qquad K_{j,P}=\bra{B_n}O_{j,P}\ket{M_n}.
\]

\paragraph*{Step 3: the two new columns.}
$K_{j,P}$ vanishes identically. Indeed, every $O_{j,P}$ applies an $X$ or a $Z$ to at least one qubit outside $E$: the strings $O_{j,Z}$, and $O_{j,Y}$ for $j<n$, end in $Z_n$; $O_{n,X}$ and $O_{n,Y}$ contain $X_{n-1}$, with $n-1\notin E$; and each $O_{j,X}=X_{j-1}X_j$ touches at least one qubit outside $E$, because no two qubits of $E$ are adjacent. On qubits outside $E$ both bra and ket carry {the state} ${\lvert}\beta{\rangle}$ or ${\lvert}\delta{\rangle}$, whose $X$ and $Z$ brackets vanish. Hence
\[
K_{j,P}=0\qquad\text{for all }j,P.
\]
$N_{j,P}$ differs from $M_{j,P}$ only on the qubits of $E$, where the factor $\bra\alpha\,\cdot\,\ket\beta$ is replaced by $\bra\alpha\,\cdot\,\ket\alpha$. By the table~\eqref{eq:brackets} these two rows agree on $I$ and $Z$ and differ on $X$ and $Y$, where $\bra\alpha\,\cdot\,\ket\alpha=0$. So $N_{j,P}=M_{j,P}$ if $O_{j,P}$ acts on $E$ only by $I$ or $Z$, and $N_{j,P}=0$ if it places an $X$ or $Y$ there. Reading off the strings: the all-$Z$ strings $O_{j,Z}$, and the strings of the boundary qubits $j=1$ and $j=n$ (whose $X$ and $Y$ factors sit on qubits $1$, $n-1$, $n$, none of which is in $E$), leave $N=M$; for $2\le j\le n-1$ the pair $\{j-1,j\}$ always contains a qubit of $E$---its even member---so $O_{j,X}$ and $O_{j,Y}$ place an $X$ or $Y$ there. Comparing with the table of $M_{j,P}$ in Appendix~\ref{app:proof},
\begin{align*}
N_{j,P}&=M_{j,P}\ \text{wherever}\ M_{j,P}=\pm(1+\ii), \\
N_{j,P}&=0\ \text{wherever}\ M_{j,P}=-1+\ii .
\end{align*}

\paragraph*{Step 4: the cosine.}
Insert these values into~\eqref{eq:threeterms}, treating the two kinds of rows separately. On rows with $N=M=s(1+\ii)$, $s=\pm1$, use $c_B+c_M=(1-\ii)e^{\ii\theta}$ and $(1-\ii)(1+\ii)=2$:
\begin{align}
\langle P_j\rangle_\tau
&=2^{-n}\,s\,\mathrm{Re}\bigl[(1-\ii e^{-\ii\theta})(1-\ii)e^{\ii\theta}(1+\ii)\bigr]\notag\\
&=2^{1-n}\,s\,\mathrm{Re}\bigl[e^{\ii\theta}-\ii\bigr]=2^{1-n}s\cos\theta .
\end{align}
On rows with $N=K=0$ and $M=-1+\ii$, use $(1-\ii)(-1+\ii)=2\ii$:
\begin{align}
\langle P_j\rangle_\tau
&=2^{-n}\,\mathrm{Re}\bigl[(1-\ii e^{-\ii\theta})\,2\ii\bigr]
=2^{1-n}\cos\theta .
\end{align}
Since $\langle P_j\rangle_0=2^{1-n}\operatorname{Im}M_{j,P}$ equals $2^{1-n}s$ and $2^{1-n}$ on the two kinds of rows, in both cases
\[
\langle P_j\rangle_\tau=\cos\theta\,\langle P_j\rangle_0 .
\]
Hence $\vec m_j(\tau)=\cos(\tfrac{\pi\tau}{2})\,\vec m_j(0)$ and $r(\tau)=r_{\mathrm{EJM}}\lvert\cos(\tfrac{\pi\tau}{2})\rvert$; at $\tau=1$ this is $0$. \hfill$\qed$

\paragraph*{Remark: the support is forced.}
The proof used exactly two properties of the set $E$: no two of its qubits are adjacent (Step~3, $K=0$), and it contains one qubit of every pair $\{j-1,j\}$ with $2\le j\le n-1$, while the qubits $1$, $n-1$ and $n$ stay outside it (Step~3, pattern of $N$). Walking up from the first qubit, these two requirements leave no freedom: the pair $\{1,2\}$ forces $2\in E$, adjacency then excludes $3$, the pair $\{3,4\}$ forces $4\in E$, and so on, so $E=\{2,4,\dots\}$; the last pair $\{n-2,n-1\}$ then requires $n-2\in E$, which fits this pattern only when $n$ is even. For odd $n$ no set satisfies both requirements---this is why the family is a feature of even qubit number. The numerical scan quoted in Sec.~\ref{sec:even} confirms that, beyond this proof mechanism, no other support of the form~\eqref{eq:phi} yields the cosine law at all.

\section{A second four-qubit line}\label{app:159}

This appendix records a numerical observation about the four-qubit manifold; it plays no role in the proofs. In an auxiliary four-qubit enumeration, analogous in spirit to the three-qubit search of Ref.~\cite{MultiqubitEJM}, we found a second regular tetrahedral solution, with the larger Bloch length $3\sqrt3/8$. (The label ``Class~159'' appearing in the accompanying code is just its index in that enumeration, which grouped candidates by simple local-unitary invariants without claiming a complete classification.) It is the modulo-$4$ phase polynomial
\begin{align}
f_{159}={}&z_2z_3+3z_3z_4+2z_1z_2z_3+z_1z_2z_4\notag\\
&+3z_1z_3z_4+z_1z_2z_3z_4 ,
\end{align}
which we identify with its table of values in $\{0,1,2,3\}$ on the strings $0000,0001,\dots,1111$,
\begin{equation}
f_{159}=(0,0,0,3,\,0,0,1,0,\,0,0,0,2,\,0,1,3,3).
\end{equation}
Numerically, this solution also lies on a straight line of real phase functions ending at a $1$-uniform point: with the endpoint table
\begin{align}
g_{159}={}&(0,\tfrac23,-\tfrac23,\tfrac73,\,0,\tfrac23,1,0,\notag\\
&-\tfrac23,\tfrac43,0,\tfrac83,2,3,\tfrac{11}{3},3),
\end{align}
the segment
\begin{equation}
f^{(159)}_\tau=(1-\tau)\,g_{159}+\tau f_{159},\qquad 0\le\tau\le1,
\end{equation}
stays regular tetrahedral to machine precision; at $\tau=0$ all one-qubit Bloch vectors vanish, and at $\tau=1$ the Bloch length is $3\sqrt3/8$. (Adding multiples of $4$ to entries of $f_{159}$ leaves its gate unchanged but tilts the straight line, so the segment is defined with the reduced values above.)

For comparison, the even-family endpoint at $n=4$ is a product of Bell pairs on $(1,2)$ and $(3,4)$. The endpoint above is not of that form: across the cut $(12)|(34)$ its Schmidt coefficients are approximately
\[
(0.583,0.483,0.470,0.454),
\]
which have full rank. Thus the four-qubit manifold appears to contain different inequivalent $1$-uniform endpoints. Whether the corresponding branches can be joined globally remains open.

\begin{figure}[t]
\centering
\includegraphics[width=0.95\columnwidth]{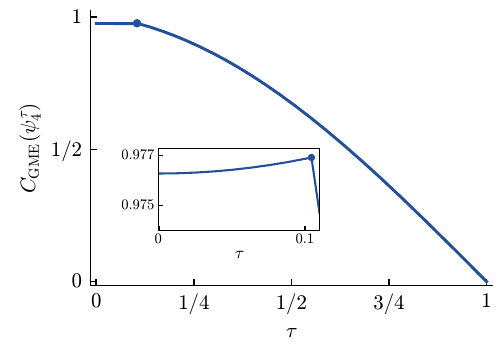}

\caption{Genuine multipartite concurrence of the four-qubit family, Eq.~\eqref{eq:gme-four}. The inset resolves the shallow maximum at $\tau\simeq0.1044$; the concurrence vanishes at the Bell-pair endpoint $\tau=1$.}
\label{fig:gme-concurrence}
\end{figure}

\section{Multipartite entanglement}\label{app:multipartite}
The single-qubit entropies do not characterize all multipartite entanglement. Three-tangle and pairwise concurrence for the three-qubit EJMs were already studied in Ref.~\cite{MultiqubitEJM}, whose End Matter (Table~I) includes the present class:
\[
\mathcal T_3=\frac{\sqrt{97}}{16},\qquad C_{AB}^2=C_{AC}^2=C_{BC}^2=\frac{13-\sqrt{97}}{32}.
\]
These values, like all local-unitary invariants, are shared by every outcome of the Pauli-orbit basis.

For the four-qubit family, the pure-state genuine multipartite concurrence~\cite{Ma2011GME},
\begin{equation}\label{eq:gme-definition}
C_{\mathrm{GME}}(\psi)=\min_{A|\bar A}\sqrt{2\bigl(1-\operatorname{Tr}\rho_A^2\bigr)},
\end{equation}
where the minimum runs over all nontrivial bipartitions, is
\begin{equation}\label{eq:gme-four}
C_{\mathrm{GME}}(\psi_4^\tau)=\frac18\sqrt{\min\!\left\{61+3u^2,(1-u)(65+49u)\right\}},
\end{equation}
with $u=\sin(\pi\tau/2)$. It is positive for $\tau<1$ and vanishes at the Bell-pair endpoint. Thus maximal entanglement across every single-qubit cut, or $1$-uniformity, need not imply genuine four-party entanglement. The same expression holds for every measurement outcome.

Equation~\eqref{eq:gme-four} can be checked directly in SymPy by evaluating the reduced-state purities for the seven inequivalent bipartitions. Its dependence on $\tau$ is shown in Fig.~\ref{fig:gme-concurrence}.

\section{A square analogue}\label{sec:square}

The same construction yields other interesting geometries. Subtracting the top symmetric function,
\begin{equation}\label{eq:square}
f_\square=f_{\mathrm{EJM}}-e_n,
\end{equation}
shifts only the phase of the all-ones string by $-\pi/2$ and flattens each local tetrahedron into a square. For every $n\ge3$ the orbit's one-qubit Bloch vectors form congruent squares, each with one vanishing Bloch component and common radius $r_\square=\sqrt2/2^{\,n-2}$; the explicit Bloch vectors are
\begin{equation}
\vec m_j^\square=2^{2-n}\!\begin{cases}(1,1,0)&j<n,\ n-j\ \text{odd},\\(1,0,1)&j<n,\ n-j\ \text{even},\\(0,-1,1)&j=n.\end{cases}
\end{equation}
The formula follows directly from the EJM computation.  Indeed,
$f_\square=f_{\rm EJM}-e_n$ only changes the phase of the all-ones
computational-basis string.  Thus, in each one-qubit Bloch component, all EJM
terms are unchanged except the single pair involving that string;
checking this pair turns one EJM component to zero and leaves the other two
equal.

Planar geometries of this kind were already constructed in Ref.~\cite{Pauwels2026PRA}, which gives a real-valued fiducial whose Bloch vectors form rectangles in the $X$--$Z$ plane for every $n$. At $n=3$ a direct calculation shows that the two constructions share squares of radius $\sqrt2/2$, identical Schmidt spectra across every cut, and vanishing three-tangle, but they differ beyond it: the rectangles of Ref.~\cite{Pauwels2026PRA} are square only at $n=3$ and grow with the number of qubits, whereas~\eqref{eq:square} keeps a perfect square on every qubit for every $n$, shrinking at the same rate as the EJM tetrahedron. Both are members of a wider class of square and rectangular orbit bases (already at $n=3$ there are inequivalent squares), whose classification we leave open.

\end{document}